\documentclass[12pt]{article}

\usepackage{amsmath,amsfonts,amssymb,theorem,verbatim,graphics,epsfig}
\numberwithin{equation}{section}
\newtheorem{Thm}{Theorem}[section]
\newtheorem{Cor}{Corollary}[section]
\newtheorem{Lemma}{Lemma}[section]
\newtheorem{Prop}{Proposition}[section] 
{\theorembodyfont{\normalfont}
\newtheorem{Def}{Definition}[section]

}

\def\BEN{\begin{enumerate}}  \def\BI{\begin{itemize}}
\def\EEN{\end{enumerate}}   \def\EI{\end{itemize}} 
   
  \def\no{\noindent}

\def\nn{\nonumber}
\def\beq{\begin{eqnarray}} \def\eeq{\end{eqnarray}}

\def\al*#1{\begin{align*}#1\end{align*}}

\def\ga*#1{\begin{gather*}#1\end{gather*}}

\def\alat*#1#2{\begin{alignat*}{#1}#2\end{alignat*}}
\def\bea{\begin{eqnarray*}}
\def\eea{\end{eqnarray*}}
\def\ml*#1{\begin{multline*}#1\end{multline*}}

 \def\mbf{\mathbf} \def\mrm{\mathrm}

\newcommand{\Prob}{{\rm I\hspace{-0.7mm}P}}
\newcommand{\Exp}{{\rm I\hspace{-0.7mm}E}}

  \def\R{{\mathbb R}}

\def\mc{\mathcal}

\def\le{\left} \def\ri{\right} \def\i{\infty}

\def\te#1{\mathrm{e}^{#1}}  \def\td{\text{\rm d}}
 
\def\I{\int} 

\def\T{\tilde}  
 \def\WT{\widetilde}

 \def\b{\beta}
  \def\d{\delta}   \def\th{\theta}

\def\e{\epsilon}  \def\l{\lambda}  
  \def\nn{\nonumber}   \def\s{\sigma}
\def\t{\tau}     
\def\c{\chi} \def\w{\omega} \def\q{\qquad} \def\D{\Delta}
  \def\L{\Lambda}

\newcommand{\proof}{\no{\it Proof\ }}

\newcommand{\exit}{{\mbox{\, \vspace{3mm}}} \hfill\mbox{$\square$}}

\begin{document}
\addtocounter{page}{-1}
\title{Optimal dividend distribution \\ under Markov-regime switching%
\footnote{Research supported by EPSRC grant EP/D039053. The
research was carried out in part while the authors were based at
King's College London.
\newline
This paper is based on Chapter 3 of the first author's PhD thesis
(submitted in April 2008).
\newline
{\sl Acknowledgements:} We thank A. Dassios, R. Norberg, M.
Zervos, participants of the 12th International Congress of IME
(Insurance: Mathematics and Economics) in Dalian, China, and the
Bachelier Conference in London (July 2008) for useful suggestions.
We are grateful to
Chengming Xu and two anonymous referees for helpful comments and a careful reading of the paper.
}}
\author{Zhengjun Jiang\footnote{Statistics Programme, Division of Science and Technology,
Beijing Normal University-Hong Kong Baptist University, United
International College, Zhuhai, P. R. China 519085,
 email: {\tt zjunjiang@yahoo.com.cn}, Tel: +86 756 3620624, Fax: +86 756
 3620888}
\ \&\ Martijn Pistorius\footnote{Department of Mathematics,
Imperial College London, London SW7 2AZ, UK, email: {\tt
m.pistorius@imperial.ac.uk}, Tel: +44 20 7594 8532, Fax: +44 20
7594 8483}
\\
\\
}
\date{To appear in {\sl Finance and Stochastics}} \maketitle
\thispagestyle{empty}

\begin{abstract}
We investigate the problem of optimal dividend distribution for a
company in the presence of regime shifts. We consider a company
whose cumulative net revenues evolve as a Brownian motion with
positive drift that is modulated by a finite state Markov chain,
and model the discount rate as a deterministic function of the
current state of the chain. In this setting the objective of the
company is to maximize the expected cumulative discounted dividend
payments until the moment of bankruptcy, which is taken to be the
first time that the cash reserves (the cumulative net revenues
minus cumulative dividend payments) are zero. We show that, if the
drift is positive in each state, it is optimal to adopt a barrier
strategy at certain positive regime-dependent levels, and
provide an explicit characterization of
the value function as the fixed point of a
contraction.  In the case that the drift is small and negative in
one state, the optimal strategy takes a different form, which we
explicitly identify if there are two regimes. We also provide a
numerical illustration of the sensitivities of the optimal
barriers and the influence of regime-switching.

\medskip


\medskip


\medskip


\end{abstract}

\newpage

\section{Introduction}
A classical topic in finance and actuarial science is that of
optimal dividend distribution for a company, which can be phrased
as the problem of determining the optimal timing and sizes of
dividend payments in the presence of bankruptcy risk, where the
usual objective is to maximize the expected value of the
cumulative discounted dividend payments until bankruptcy. The
earliest work in this setting can be traced back to De Finetti
\cite{de:fin} who studied the dividend problem for an insurance
company under the binomial model. In continuous time the problem
was posed and solved in a Brownian motion model for the cash
reserves by Jeanblanc-Piqu\'{e} and Shiryaev \cite{JeanShir}, and
Asmussen and Taksar \cite{Asm:Tak}, using optimal control theory.
Since then an extensive literature has appeared on the dividend
problem and its extensions, including reinsurance (e.g.
\cite{Schm}), optimal investment of the reserves (e.g.
\cite{H:taks2}), tax and proportional cost (e.g.
\cite{Cad:Sak:Zap,LZ}), and growth options (\cite{Dec:Vil}).
%

In general, the form of the optimal dividend policy has been found
to depend on the expected growth rate and variability of future
revenues, and the discount rate. These quantities will evolve in
time reflecting changing market and economic conditions, and those
changes may happen gradually or occur abruptly and be more
substantial. Here we will focus on the changes of the latter type
(also called regime shifts or switches) and model the cumulative
net revenues of the company as a Brownian motion with the drift
and volatility modulated by a finite state Markov chain, and the
discount rate as a deterministic function of the chain.
Since Hamilton \cite {hamilton, hamilton2}, a substantial
econometric literature has appeared that supports the use of
Markov regime-switching models to describe business cycles, term
structure of interest rates and other macroeconomic quantities.
Such models have been shown to be capable of capturing occasional
simultaneous and substantial changes of the parameters.
Regime-switching models also have the advantage of retaining a
degree of analytical tractability, and models from this class can
in principle approximate a given diffusion arbitrarily closely by
taking the state space large enough and specifying the generator
matrix appropriately. In the mathematical finance literature
regime-switching models have become more popular, and have found
their applications in stock price models, interest rate models and
the real option literature. See e.g. Boyarchenko and
Levendorski\v{i} \cite {boy:leve}, Buffington and Elliott
\cite{buf:ell}, Driffill et al. \cite {driffi}, Duan et al. \cite
{duan:et}, Elliott et al. \cite {ell2}, Guo and Zhang \cite
{GuoZ}, Jiang and Pistorius \cite{MP1}, Naik \cite {naik} for
derivative pricing, Elliott and van der Hoek \cite {ell:van} and
Guidolin and Timmermann \cite {gui:tim} for asset allocation,
B{\"a}uerle \cite {bau}, Li and Lu \cite {li:lu}, Zhu and Yang
\cite {zhu:ya} and Asmussen \cite{asmu} for ruin and risk theory,
and Guo et al. \cite {GMM} for irreversible investment.

In this regime-switching setting we will consider the problem of
the management of the company to find a dividend distribution
policy that maximizes expected discounted dividend payments until
bankruptcy, which is defined to occur at the first moment when the
level of the cash reserves hits zero. We will restrict ourselves
to the case that the management can only control timing and size
of the dividend payments. In the case that the drift is positive
in every regime, we will show that it is optimal to adopt a
barrier-type strategy at certain positive levels that depend on
the current regime, that is, it is optimal to make the minimal
payments needed to keep the cash reserves below these barrier
levels. When a regime-switch occurs, dividend payments are to be
postponed or brought forward in time, according to whether the
barrier jumps up or down, and in the latter case a lump sum should
be paid if the reserves were above the new barrier at the moment
of the switch. In the case of a single regime this strategy
reduces to the classical constant barrier strategy that was found
before by Asmussen and Taksar \cite{Asm:Tak}.

After an adverse economic regime-switch it could happen that the
expected net revenue of the company becomes negative, in which
case the optimal strategy takes a different form. Intuitively, it
is clear that, if the drift is negative and the reserves are
sufficiently small, it will be optimal to liquidate the company by
paying out the reserves as a lump-sum. In the absence of
regime-switching, this optimality actually holds irrespective of
the size of the reserves. In the presence of regime-switching,
however, we find that it is optimal to continue the business if
the drift is small and negative and the reserves are not too
small: the prospect of switching to a better regime with suitable
positive drift outweighs the risk of ruin. In this case the value
function is not concave, which differs from what is usually
found in singular control problems. An explicit solution is
derived in Section \ref{sec:casetwo} in the case of two regimes.

The dividend optimization problem gives rise to a singular control
problem, whose HJB equation takes the form of a coupled system of
variational inequalities, due to the fact that the problem is
driven by a two-dimensional Markov process. A commonly used direct
approach for explicitly solving optimal control problems proceeds
by guessing a candidate optimal solution, constructing a
corresponding value function, assuming smoothness if necessary,
and subsequently verifying its optimality by employing a
verification result. Here we shall follow a different approach to
construct the candidate value function, by directly employing a
dynamic programming equation. We will prove that the value
function is the fixed point of a certain contraction operator,
which is given explicitly in terms of the initial data, and derive
an explicit iterative algorithm to calculate the value function,
which `decouples' the different regimes such that at any stage
one-dimensional control problems are solved. This construction
yields in particular that the value function is $C^2$, which
implies that the value function is a classical solution of the HJB
equation. At this point it is worth mentioning that, although it
is possible to follow the direct approach, this seems to become
intractable if the number of states is large, as it leads to a
large collection of systems of coupled non-linear equations
(corresponding to different orderings of the dividend levels).

After the first version of this paper was written, we discovered a
related work on optimal dividend problems by Sotomayor and
Cadenillas \cite{CaSo}. In a setting that is a particular case of
ours, with two regimes and constant rate of discounting, they
solve three dividend distribution problems with bounded and
unbounded dividend rates, and in the presence of fixed cost,
respectively, under the assumption of existence of
a solution to the smooth fit equation.

The remainder of the paper is organized as follows. In Section
\ref{sec:prel} we give a statement of the problem, and present a
dynamic programming equation and related theorem. In Sections
\ref{sec:opt} and \ref{sec:alg} we present the optimal solution
and give a proof by constructing an iterative algorithm to
calculate the value function $V$. Section \ref{sec:casetwo} is
devoted to a case study of the setting of two regimes, with a
numerical illustration of the sensitivities of the optimal barrier
levels to the different parameters. Section \ref{sec:concl}
concludes. Some proofs are presented in the Appendix.

\bigskip
\section{Preliminaries and first results}\label{sec:prel}
\subsection{Problem formulation}

Let $\{W_t: t\ge 0\}$ be a Wiener process and let $\{Z_t: t\ge
0\}$ be a continuous time Markov chain with finite state space $E$
and generator matrix $Q=(q_{ij})_{i,j\in E}$, independent of $W$.
Assume that the cash reserves $X=\{X_t,t\ge0\}$ evolve, in the
absence of dividend payments, as a regime-switching
linear Brownian motion, that is, $X$ satisfies the SDE
$$
\td X_t =  \mu(Z_t)\td t + \sigma (Z_t)\td W_t, \qquad
X_0=x>0,\quad Z_0=i,
$$
where $Z$ represents the state of economy. For every state $i$ in
$E$, both drift parameter $\mu(i)$ and volatility parameter
$\sigma(i)>0$ are assumed to be known constants. In case there is
no notational confusion possible, we will write $\mu_i$ and
$\sigma_i$ for $\mu (i)$ and $\sigma (i)$ respectively. The
processes $X$ and $Z$ are defined on some filtered probability
space $(\Omega, \mathcal F, \mathbf F, \Prob)$ where $\mathbf F =
\{\mathcal F_t, t\geq 0\}$ denotes the right-continuous completed
filtration jointly generated by $X$ and $Z$. We denote by
$\Prob_{x,i}$ and $\Prob_x$ the measure $\Prob$ conditioned on
$\{X_0=x, Z_0=i\}$ and $\{X_0=x\}$, respectively, and write
$\Exp_{x,i}$ and $\Exp_x$ for the corresponding expectations. We
assume that the processes $X$ and $Z$ are both fully observable to
the shareholders, and that these decide on the dividend strategies
on the basis of the available information.

A dividend strategy $D$ is a non-decreasing and right-continuous
stochastic process  $D=\{D_t: t\ge0\}$ with $D_{0-}=0$. Here $D_t$
represents the cumulative amount of dividends that has been paid
out until time $t$. We will assume that, apart from reducing the
reserves, dividend payments have no effect on the business and
that there are no transaction costs associated to the payment or
receipt of dividends.
The dynamics of the risk reserve process $U=\{U_t: t\ge0\}$ in the
presence of dividend payments are then given by
\begin{equation}\label{eq:U}
\td U_t = \td X_t - \td D_t
\end{equation}
for all $t$ until the time $\tau$ of bankruptcy and $\td U_t=0$
for $t$ after $\tau$, where
\begin {equation*}
\tau=\inf \{t\geq 0:\,U_t = 0 \}
\end {equation*}
is the first time that $U$ hits zero. To avoid degeneracies only
those dividend strategies will be considered that have no lump sum
dividend payments larger than the current level of the reserves:
A dividend strategy $D$ is called {\em admissible} if $D$ is
$\mathbf F$-adapted, $\td D_t = 0$ for $t\ge\tau$ and
\begin{equation}\label{eq:ls}
U_{t-}\geq D_{t}-D_{t-}\q \text{for all $t<\tau$.}
\end{equation}
 Denoting by $\mathcal D$ the set of admissible dividend
strategies, the objective function of the shareholders is given by
\begin {equation} \label {value:f}
V(x,i) = \sup_{D\in\mathcal D} V_D(x,i),
\end {equation}
where $V_D$ denotes the expected value of the discounted dividends
until the time of ruin $\tau$ under the dividend strategy $D$,
$$
V_D(x,i) = \Exp_{x,i} \le[\I_0^\tau \te{-\I_0^t r(Z_s)\td s}\td
D_t\ri],
$$
with $r:E\to(0,\infty)$ the Markov-modulated rate of discounting.
The problem for the shareholders is to identify a dividend
strategy $D^*\in\mathcal D$ that attains the supremum in
\eqref{value:f}, that is, $V \equiv V_{D^*}$.

\subsection{A priori bounds}
Assume for the moment that there is only a single regime,
$E=\{i\}$. Then we are back in the classical linear Brownian
motion setting that was investigated in Asmussen and Taksar
\cite{Asm:Tak}. They showed that, if $\mu_i>0$, the optimal
strategy is a {constant barrier strategy} at the level
\begin {equation}\label{noregO}
a_i^*=\frac {\sigma_i^2}{\sqrt{\mu_i^2+2r_i\sigma_i^2}}\ln \left(
\frac
{\sqrt{\mu_i^2+2r_i\sigma_i^2}+\mu_i}{\sqrt{\mu_i^2+2r_i\sigma_i^2}-\mu_i}
\right).
\end {equation}
According to this strategy, the overflow of the reserves
above the level $a_i^*$ is immediately paid out as dividends.
The corresponding value function is given by
\begin {equation}\label {noregV}
V^*_i(x)=
\begin {cases}
W^{(r_i)}_i(x)/W^{(r_i)\prime}_i(a_i^*)
\,,&0\leq x\leq a_i^*,\\
x-a_i^*+\mu_i/r_i, &x\geq a_i^*,\\
\end {cases}
\end {equation}
where
\begin{equation}\label{eq:Wq}
W^{(q)}_i(x) = \frac{2}{\s_i^2}\cdot\frac{\te{\l_i^+
x}-\te{\l_i^{-}x}}{\l_i^+ - \l_i^-},
\end{equation}
where $\l^-_i<0 < \l^+_i$ denote the roots of the equation $\frac
12 \sigma_i^2 \lambda^2+\mu_i \lambda-q=0$:
\begin{equation}\label{noswit:root}
\l_i^\pm = \l_i^\pm(q) = - \frac{\mu_i}{\sigma_i^2} \pm
\sqrt{\le(\frac{\mu_i}{\sigma_i^2}\ri)^2+\frac{2q}{\sigma_i^2}}.
\end{equation}
The eqs. \eqref{noregO}---\eqref{noswit:root} show that the value
function and optimal level are both functions of the drift and of the
rate of discounting per unit of squared volatility. This
observation leads one to expect that $V(x,i)$ is bounded above and
below by the values $V_{+}(x)$ and $V_{-}(x)$ of firms operating
in a more or less favourable environment, with volatility constant equal to one drift and discounting equal to
$(\frac{\mu_{+}}{\sigma^2_{+}},\frac{r_{+}}{\sigma^2_{+}}) =
(\max_{i\in E}\frac{\mu_i}{\sigma^2_i},\min_{i\in
E}\frac{r_i}{\sigma^2_i})$ and
$(\frac{\mu_{-}}{\sigma^2_-},\frac{r_-}{\sigma^2_-}) = (\min_{i\in
E}\frac{\mu_i}{\sigma^2_i}, \max_{i\in E}\frac{r_i}{\sigma^2_i})$,
respectively. The following result confirms that these explicit
bounds indeed hold true:

\begin{Prop}\label{prop:apriori}
If $\mu_->0$, it holds that
\begin{equation}\label{lub}
V_{-}(x) \leq V(x,i) \leq V_{+}(x)
\end{equation}
for all  $x\ge 0, i\in E$.
\end{Prop}
The bounds in \eqref{lub} will be employed in the construction of
the optimal value function in Section \ref{sec:alg}.

\subsection{Dynamic programming equation and comparison result}
The following dynamic programming equation for
the value function of the singular control problem \eqref{value:f}
will form the basis for its solution:

\begin{Prop}\label{prop:DP} It holds that
\begin{equation}\label{eq:DP}
  V(x,i) = \sup_{D\in\mathcal D}
  \Exp_{x,i}\le[\int_0^{\tau\wedge\zeta}\te{-\Lambda_{t}}\td D_t
  + \te{-\Lambda_{\tau\wedge\zeta}}V(U_{\tau\wedge\zeta},
  Z_{\tau\wedge\zeta})\ri],
\end{equation}
where $\zeta$ denotes the epoch of the first regime-switch and
$\Lambda_t = \int_0^t r(Z_s)\td s$.
\end{Prop}
The proof of Proposition \ref{prop:DP} is given in the Appendix.
This dynamic programming equation is associated with the following
Hamilton-Jacobi-Bellman equation for the value function: 
\begin{equation}\label{eq:HJB}
\max\le\{\mathcal G w(x,i) - r(i) w(x,i), 1 - w'(x,i)\ri\}=0,\q
x>0,\,i\in E,
\end{equation}
where $\prime$ denotes the partial derivative with respect to $x$
and $\mathcal G$ denotes the infinitesimal generator of $(X,Z)$
which acts on functions $w: [0,\infty) \times E \to [0,\infty)$
with $w(\cdot, i)\in C^2([0,\i))$ for $i\in E$ as $\mathcal G
w(x,i) = \mathcal G_o w(x,i) + \mathcal G_s w(x,i)$ where
\begin {equation}\label {operator:value}
\mathcal G_o w(x,i) = \frac {\s_i^2}{2}w''(x,i) + \mu_i w'(x,i),
\quad \mathcal G_s w(x,i) = \sum_{j\in E}q_{ij}[w(x,j) - w(x,i)].
\end {equation}
It holds that any sufficiently regular super-solution of the
 HJB equation \eqref{eq:HJB} dominates the value function:

\begin{Thm}\label{verif:thm}
Assume that there exists a function $w=(w(\cdot,i), i\in E)$, with
$w(\cdot,i)$, $i\in E$, $C^1$ functions on $(0,\i)$ that are
piecewise $C^2$ and satisfy for $x>0$
\begin{eqnarray*}
\mathcal G w(x,i) - r(i) w(x,i) &\leq& 0 \q \text{in distributional sense},\\
w(0,i) = 0,\quad w'(x,i) &\ge& 1.
\end{eqnarray*}
(i) Then it holds that $w(x,i)\ge V(x,i)$ for all $x\geq 0$ and
$i\in E$.

\no (ii) If, in addition, $w=V_D$ for some $D\in\mathcal D$, then
$D$ is an optimal strategy and $V\equiv w$.
\end{Thm}
\proof{\it of Theorem \ref{verif:thm}.} (i) Fix an arbitrary
$D\in\mathcal D$ and let $U$ be the corresponding risk process.
The statement will follow once we have shown that $w(x,i)\ge
V_D(x,i)$. Applying a generalised form of It\^{o}'s lemma to the
process $\{\te{-\Lambda_{T\wedge\tau}}w(U_{T\wedge \tau},
Z_{T\wedge \tau}), T\ge 0\}$, we find that
\begin{eqnarray}\label{eq:discount}
\nn
\lefteqn{\te{-\Lambda_{T\wedge\tau}}w(U_{T\wedge\tau},Z_{T\wedge\tau})
- w(U_0,Z_0) + \I_{0}^{T\wedge\tau} \te{-\Lambda_{s}}\td D_s}\\
\nn &=& \I_{0}^{T\wedge\tau} \te{-\Lambda_{s}}  (\mc Gw - rw)
(U_{s-}, Z_{s})  \td s + \I_{0}^{T\wedge\tau}
\te{-\Lambda_{s}}\le[1 -
w'(U_{s-}, Z_{s-})\ri]\td D^c_s \nonumber \\
&\phantom{=}&+ \sum_{0\leq s\leq
T\wedge\tau}\te{-\L_s}\le[w(U_{s-} - \D D_s, Z_s) - w(U_{s-},
Z_{s-}) +\D D_s\ri]1_{\{\D D_s > 0\}} + M_{T\wedge \tau}
\end{eqnarray}
where $D_t = D^c_t + \sum_{0\leq s\leq t}\Delta D_s$ and
$M_{T\wedge\tau}$ is the local martingale with
$$
M_{t}=\I_{0}^{t}\te{-\Lambda_{s}}\sigma(Z_{s-})w'(U_{s-},
Z_{s-})\td W_s
    + \I\te{-\Lambda_{s}}\le[w(U_{s-},j) -
w(U_{s-}, Z_{s-})\ri]\WT\pi(\td s,\td j).
$$
Here the last integration is over the set ${[0,t]\times E}$
and $\WT\pi = \pi-\nu$ is a compensated random
measure\footnote{see e.g. Jacod and Shiryaev~\cite[II.1.16]{JS}
for background on random measures} where $\pi(\td t,\td j)=
\sum_{s\ge 0}1_{\{\Delta Z_s(\w)\neq 0\}} \d_{(s,Z_s(\w))}(\td
t,\td j)$, with $\d_{(s,z)}$ denoting the Dirac measure at point
$(s,z)$, and the compensator $\nu$ is given by
$$
\nu(\td t,\td j) = p_{Z_{t-}}(j) [-q_{Z_{t-},Z_{t-}}]\;\delta(\td
j) \td t = q_{Z_{t-},j}\;\delta(\td j) \td t, \quad j\in E,
$$
where $p_{Z_{t-}}(j) = \frac{q_{Z_{t-},j}}{-q_{Z_{t-},Z_{t-}}} =
P(Z_t=j|Z_{t-},\D Z_t\neq 0)$, where $\delta$ is the counting measure on $E$. Notice from \eqref{eq:discount}
that, as $M$ is bounded below and $M_0=0$, $M$ is a
super-martingale with $\Exp[M_{T\wedge\tau}]\leq 0$. In view of
HJB equation~\eqref{eq:HJB}, the right-hand side
of~\eqref{eq:discount} is non-positive, so that taking
expectations yields that
$$
w(x,i) \geq  \Exp_{x,i} \left [
\te{-\Lambda_{T\wedge\tau}}w(U_{T\wedge\tau},Z_{T\wedge\tau})\right]
+\Exp_{x,i} \left [ \int_0^{T\wedge \tau} \te{-\Lambda_s} \,
\mathrm d D_s \right ].
$$
By letting $T\to\infty$ and invoking the monotone convergence
theorem and the fact that $w$ is non-negative, we obtain that
$w(x,i) \geq  V_D(x,i)$ and hence $w(x,i) \geq V(x,i)$. (ii) The
equality follows since $V_D\leq V$ (by definition of $V$) and
$V_D\ge V$ (by part (i)).\exit


\section{The optimal dividend strategy}\label{sec:opt}
Following the classical approach to solving optimal control
problems we next construct a candidate optimal solution. In view
of the fact that $(U,Z)$ is a Markov process we consider
strategies that pay out the overflow of the cash reserves above a
regime-dependent level:

\begin{Def} A {\em modulated barrier strategy} at level
$b=(b(i), i\in E)$ is a dividend strategy $D^b\in\mathcal D$
satisfying
\begin{eqnarray*}
&\mrm{(i)}&\quad  \int_0^\i 1_{\{U_t^b < b(Z_t)\}}\td D^b_t = 0, \\
&\mrm{(ii)}&\quad U^b_t \leq b(Z_t)\ \text{for any $t\ge 0$},
\end{eqnarray*}
where $U^b$ is the risk-process \eqref{eq:U} corresponding to
$D^b$.
\end{Def}

\begin{figure}[tp]
\centering
        \includegraphics[width=10cm]{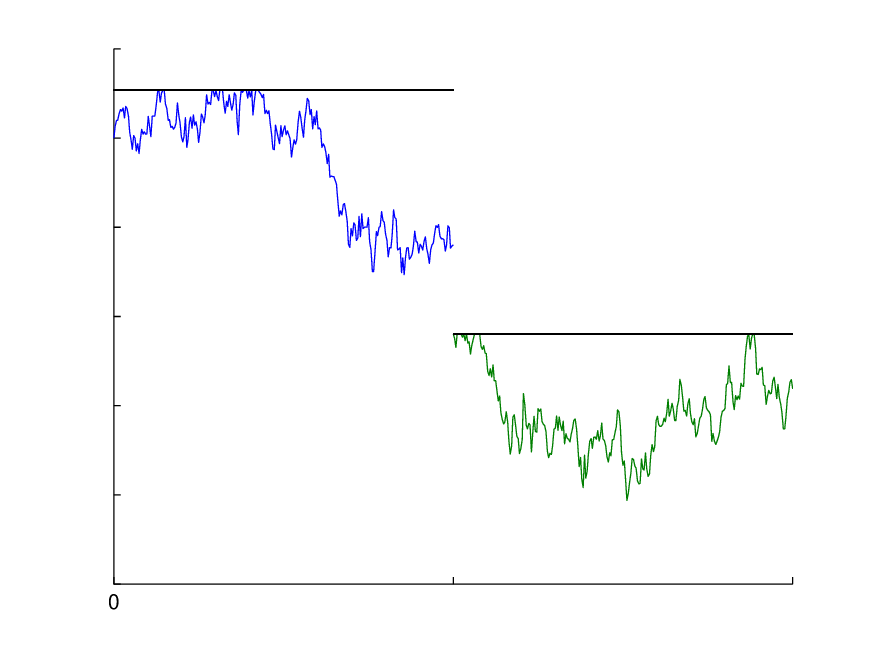}
\caption{ {\small Illustrated is the cash reserves process
corresponding to a modulated barrier strategy. The barrier levels
are represented by horizontal lines. In this case the barrier
jumps down at the moment of the regime-switch and a lump sum
payment is made.} } \label{fig:VG}
\end{figure}

According to this strategy, dividends are only paid out when
$U^b$ is at the barrier $b$, which implies that process $D^b$ is a local
time. It is straightforward to verify that $D^b$ can be explicitly
 expressed in terms of a running supremum as follows:
\begin{equation}\label{eq:Lbt}
D^b_t = 0 \vee \sup\limits_{0\leq s\leq
t}{\left\{x+\int_0^s\mu(Z_u)\td u+\int_0^s\sigma (Z_u)\td W_u -
b(Z_s) \right\}}.
\end{equation}
Employing the heuristic `principle of $C^2$ fit' of singular
control allows us to define candidate optimal levels as the
solution of the system of equations
\begin{equation}\label{eq:smoothfit}
V^{b\prime\prime}_i(b_i) = 0, \qquad i\in E,
\end{equation}
if such a solution exists. In fact, \eqref{eq:smoothfit} follows
from Lemma \ref{lem:max} and Proposition \ref{prop:fp} as you can
see later. If the drift is positive in all regimes, this candidate
solution is indeed optimal:

\begin{Thm}\label{1equiv:thm}
Suppose that $\mu_i>0$ for all $i\in E$. Then there exist levels
$b^*= (b^*_i, i\in E)$ that solve the system \eqref{eq:smoothfit},
with $0<b^*_i<\i$ and the following holds true:

(i) The optimal value function $V$  is a classical solution of
the HJB equation \eqref{eq:HJB}. In particular, $V$ is equal to
the unique solution $w=\{w(x,i),i\in E\}$ with $w(\cdot,i)\in
C^2([0,\infty))$ of the system
\begin {equation}\label{2s:eq}
\begin {cases}
 \frac 12
{\sigma_i}^2w^{\prime\prime}_i(x) + \mu_iw^\prime_i
(x)-(r_i - q_{ii})w_i(x) = -\sum_{j\neq i}q_{ij}w_{j}(x),        & 0<x<b^*_i, \\
w_i(x) = x-b^*_i+w_i(b^*_i),                          & x\ge b^*_i,\\
w_i(0) =0, \ \ w'_i(b^*_i)=1,                   & \\
\end {cases}
\end {equation}
for $i\in E$, where $w_i(x) = w(x,i)$.

(ii) The {\em modulated barrier strategy at} $b^*$
is an optimal policy in \eqref{value:f}.

\end{Thm}

If the drift condition is not satisfied, it is not
necessarily optimal to adopt a modulated barrier strategy.
Indeed, in Section \ref{sec:casetwo} we
show that in the case of two regimes with a small and negative drift
in one state and a positive in the other, the optimal
dividend barrier depends on the regime as well as on the level of
the reserves. In the following section we will give a proof
of Theorem \ref{1equiv:thm} by
presenting an iterative construction of the optimal value
function.

\section{Algorithm to compute the value function
$V$}\label{sec:alg} Throughout this section we will assume that
$\mu_i>0$ for all $i\in E$. We start by observing that the value
function $V^b$ of a modulated barrier strategy at level
$b=(b_i,i\in E)$ solves the following fixed point equation in
terms of the function $W_i^{(q)}$:

\begin{Prop}\label{prop:fp} For $i\in E$ it holds that $$V^b=T_b(V^b),$$ where, for any
$f:[0,\i)\times E\to [0,\i)$,
\begin{equation}\label{eq:Vb}
T_b(f)(x,i) =
\begin{cases}
 W^{(\theta_i)}_i(x) A^f_i(b_i) -
 \th_i^{-1}\displaystyle\sum_{j\neq i}q_{ij}\int_0^x W^{(\theta_i)}_i(x-y)f(y,j)\td y,
 & x\in[0,b_i],\\
 x - b_i + T_b(f)(b_i,i), & x\ge b_i,
\end{cases}
\end{equation}
where $\theta_i = r_i - q_{ii}$ and,
for any function $f:[0,\i)\times E\to\R$,
$A_i^f$ is given by
\begin{equation}\label{eq:Af}
A^f_i(y) = \frac{1}{W^{(\theta_i)\prime}_i(y)}\le[1 +
\th_i^{-1}\sum_{j\neq i} q_{ij}\int_0^y
W^{(\theta_i)\prime}_i(y-z)f(z,j)\td z\ri].
\end{equation}
\end{Prop}
The previous result can be utilized to calculate the value
function $V^b$ of the barrier strategy at $b$ by iterating the map
$T_b:v\mapsto T_bv$. Denote by $\mc B$ the set
$$\mc B =
\{f: f_i\in C([0,\i)),\ \text{$f_i(0)=0$, $f_i(x)/(1+|x|)$ is
bounded}\},$$ where $f_i = f(\cdot,i)$, and let $\|f\| =
\max_{i\in E}\sup_{x\ge0}\frac{|f_i(x)|}{1+|x|}$ for $f\in\mathcal
B$.

\begin{Cor}\label{cor:ct}
The map $T_b$ is a contraction on $\mc B$ with respect to the norm
$\|\cdot\|$. In particular, for $f\in\mathcal B$ it holds that
\begin{equation}\label{eq:convTn}
V^b(x,i) = \lim_{n\to\i} T^n_b(f)(x,i),
\end{equation}
where the convergence is in $\|\cdot\|$-norm
and $T^n_b(f) = T_b(T^{n-1}_b(f))$ for $n>1$ with $T^1_b=T_b$.
\end{Cor}
\proof{\it of Proposition \ref{prop:fp}.} Denote by $U^i=X^i-D^i$
the risk process corresponding to dividends $D^i$ being paid
according to a constant barrier strategy at $b_i$, with
$X^i_t=\mu_it + \sigma_i W_t$. Let $\tau^b=\inf\{t\ge 0:
U_t^b<0\}$ and $\tau^i = \inf\{t\ge0: U^i_t<0\}$ be the ruin times
of $U^b$ and $U^i$, and denote by $\zeta$ the epoch of the first
regime-switch and by $\eta(a)$ an independent exponential random
time with mean $1/a$. Then it holds that the ensemble $(U_t,
Z_0=i, t<\tau^b\wedge\zeta)$ is in distribution equal to $(U^i_t,
t<\tau^i\wedge\eta(-q_{ii}))$. Thus, the value $z_1(x,i)$ of the
discounted dividends received before $\zeta$ is given by
\begin{eqnarray*}
  z_1(x,i) &=& \Exp_{x,i}\le[\int_0^{\tau^b\wedge\zeta}\te{-\Lambda_s}\td D^b_s
  \ri]
  = \Exp_x\le[\int_0^{\tau^i}\te{-r_i s}1_{\{s<\eta(-q_{ii})\}}\td D^i_s
  \ri]\\
  &=& \Exp_x\le[\int_0^{\tau^i}\te{-(r_i-q_{ii})s}\td D^i_s \ri] =
  \frac{W^{(\theta_i)}_i(x)}{W^{(\theta_i)\prime}_i(b_i)},
\end{eqnarray*}
where $\th_i=r_i-q_{ii}$ and in the last line we used
\eqref{noregV}. Similarly, the value $z_2(x,i)$ of the discounted
dividends received after $\zeta$ satisfies, in view of the Markov
property,
\begin{eqnarray*}
  z_2(x,i) &=& \Exp_{x,i}\le[\te{-\Lambda_{\zeta\wedge\tau^i}}
  V^b(U^b_{\zeta\wedge\tau^i},Z_{\zeta\wedge\tau^i})\ri]
= \frac{-q_{ii}}{\theta_i} \Exp_x\le[V^b(U^i_{\eta(\theta_i)},
Z_{\eta(\th_i)}) 1_{\{\eta(\th_i) < \tau^i\}}\ri]\\
&=& \sum_{j\neq i} \frac{q_{ij}}{\theta_i}
\int_0^{b_i}V^b_j(y)\Prob_x(U^i_{\eta(\theta_i)}\in\td y,
\eta(\theta_i) < \tau^i).
\end{eqnarray*}
Employing the identity (see e.g. \cite[Thm. 1]{P})
\begin{equation}\label{eq:res}
\Prob_x(U^i_{\eta(\theta_i)}\in\td y, \eta(\theta_i) < \tau^i)/\td
y = \frac{W^{(\theta_i)}_i(x)W^{(\theta_i)\prime}_i(b-y)}{W^{(\theta_i)\prime}_i(b)}-
1_{\{x\ge y\}}W^{(\theta_i)}_i(x-y).
\end{equation}
and the fact that $V^b(x,i) = z_1(x,i) + z_2(x,i)$, we find the result as
stated.\exit

\proof{\it of Corollary \ref{cor:ct}} Note that $\mc B$ endowed
with the norm $\|\,\cdot\,\|$ is a complete metric space and that
$T$ maps $\mc B$ to itself, by definition of $T$ and the fact that
$W^{(\theta_i)}_i$ is $C^1$. Subsequently we see that
\begin{eqnarray*}
  \|T_b(f) - T_b(g)\| &\leq& \max_{i\in E, x\in[0,b_i]}\left|\sum_{j\neq i}
\frac{q_{ij}}{\theta_i} \int_0^{b_i} h_i(x,y)(f_j - g_j)(y)\td y\right|
\leq C \|f-g\|,
\end{eqnarray*}
where $h_i$ is given in \eqref{eq:res}
(with $b$ replaced by $b_i$)
and $C = \max_i\sum_{j\neq i}\frac{q_{ij}}{\theta_i}< 1$. Here we
used that $\int_0^{b_i} h_i(x,y)\td y = 1 -
\Exp_{x,i}[\te{-\theta_i\tau^i}]$. Thus, it follows that $T$ is a
contraction on $\mc B$, which implies the convergence in \eqref{eq:convTn}.\exit

\subsection{Iteration}
As next step we consider the auxiliary control
problem with a prescribed pay-off
function $v$ to be received at the epoch of the first
regime-switch $\zeta$:
\begin{equation}\label{eq:Uv}
  (Uv)(x,i) = \sup_{D\in\mathcal D}
  \Exp_{x,i}\le[\int_0^{\tau\wedge\zeta}\te{-\Lambda_t}\td D_t +
1_{\{\zeta<\tau\}}\te{-\Lambda_{\zeta}}v(U_{\zeta},
Z_{\zeta})\ri].
\end{equation}
This singular control problem can be solved explicitly if $v$ lies
in the set of smooth concave pay-off functions $\mc C =
\{v\in\mathcal B: \text{$v_i$ is increasing and concave}, i\in
E\}$:

\begin{Prop}\label{prop:Uv}
Let $v\in\mathcal C$. Then $Uv(\cdot,i)\in C^2[0,\infty)$ for $i\in E$
and the optimal strategy in \eqref{eq:Uv}
is given by a regime-switching barrier strategy at the levels
$b^v=(b^v_i, i\in E)$, $0<b^v_i<\i$, given by
\begin{equation}\label{eq:bvi}
b^v_i = \inf\{b\ge 0: A^v_i(b) \ge A^v_i(x)\ \text{for all $x\ge
0$}\},
\end{equation}
with $A^v$ given in \eqref{eq:Af}.
\end{Prop}
Supposing that the map $U: v\mapsto Uv $ preserves
concavity and smoothness, this Proposition can be applied iteratively, as follows:
Initialise by setting $n=0$ and $v=v_0$ for some
$v_0\in\mathcal B$ and then
\begin{itemize}
\item[(1)] Find $b^v=(b^v_i,i\in E)$
 in \eqref{eq:bvi}; \item[(2)]
Set $v \leftarrow T_{b^v}(v)$,\ $n\leftarrow n+1$,\ and\
$v_n\leftarrow v$, and return to step (1).
\end{itemize}
The following result shows that the sequence $(v_n)$ generated in
this way converges to the value function  $V$ as $n\to\i$:

\begin{Prop}\label{prop:iter} Let $v_0^\pm\in \mathcal C$ and define
$v_n^\pm = Uv^\pm_{n-1}$ for $n\ge 1$. If $v_0^-\leq V\leq v_0^+$, then
$v_n^- \leq V\leq v_n^+$ and
\begin{equation}\label{eq:convV}
V(x,i) =
\lim_{n\to\i} v_n^+(x,i) = \lim_{n\to\i} v_n^-(x,i),
\end{equation}
where the convergence is with respect to the norm $\|\cdot\|$.
In particular, $V$ is concave.
\end{Prop}
In fact, we shall show below that $U$ is a contraction on $\mc C$.
Notice that Theorem \ref{1equiv:thm}(i) is now a direct
consequence of these results. Indeed, by combining Proposition
\ref{prop:Uv} and the dynamic programming equation \eqref{eq:DP}
we see that the optimal strategy in \eqref{value:f} is given by a
modulated barrier strategy at some positive finite levels.
Explicit examples of initial functions $v_0^\pm$ are the $V_\pm$
given in Proposition \ref{prop:apriori}.

\subsection{Proofs}
This subsection is devoted to the proofs of the Propositions
\ref{prop:Uv} and \ref{prop:iter} which we split in a number of
steps. The first step is to verify that the $b^v_i$ as defined
above are positive and finite, which is a matter of
straightforward calculations using the explicit expression
\eqref{eq:Wq}:

\begin{Lemma}[Existence of optimal barrier levels]\label{lem:max}
Let $v\in\mathcal C$. Then $b\mapsto A^v_i(b)$ attains its maximum
at some finite and positive $b_i$, which satisfy
$T_b(v)_i''(b_i):=\frac{\partial^2}{\partial x^2}T_b(v)(x,i)|_{x=b_i} = 0$, $i\in E$. In particular, $T_{b^v}(\cdot, i)\in C^2[0,\infty))$
for $i\in E$.
\end{Lemma}

The proof of Lemma \ref{lem:max} is given in the Appendix. The key
step is to verify next that the value function of a barrier
strategy at level $b^v$ with a concave payoff function
$v(\cdot,i)$, is itself concave:

\begin{Lemma}[Preservation of concavity]\label{lem:Uv}
If $v\in\mathcal C$, then $T_{b^v}(v)\in\mathcal C$.
\end{Lemma}

\proof{\it of Lemma \ref{lem:Uv}} We first assume that $v\in\mc
C\cap C^2[0,\i)$, and write $b$ instead of $b^v$ to
simplify the notation.
In view of the smoothness of $v$ and the
definition of $w_i(x):=(T_{b^v}v)(x,i)$, we can obtain from
\eqref{eq:Wq} and \eqref{eq:Vb} that for $x\in(0,b_i)$,

$$w_i'(x)
  =W^{(\theta_i)\prime}_i(x)A_i^v(b_i)-\theta_i^{-1}\sum_{j\neq i}q_{ij}\int_0^x W^{(\theta_i)\prime}_i(x-y)v (y,j)\,\mathrm
  {d}y.
$$
and
$$
  w_i''(x)=W^{(\theta_i)\prime\prime}_i(x)A_i^v(b_i)
  -\theta_i^{-1}\sum_{j\neq i}q_{ij}\left[ W^{(\theta_i)\prime}_i(0)v (x,j)+\int_0^x W^{(\theta_i)\prime\prime}_i(x-y)v (y,j)\,\mathrm
  {d}y \right].
$$
From these expressions, equation \eqref{eq:Wq} and the $v\in
C^2[0,\infty)$, we have that $w_i|_{(0,b_i)}\in C^4(0,b_i)$. In addition,
we have $w_i'(b_i)=1$ from the above expressions and equation
\eqref{eq:Vb}, and have $w_i''(b_i)=0$ by Lemma \ref{lem:max}. As
a result, $w_i$ is $C^2[0,\infty)$.

An application of It\^{o}'s lemma shows that $w_i$ satisfies the ode
\begin{equation}\label{eq:ode}
f_i^v(x):=
\frac{\sigma_i^2}{2} w''_i(x) + \mu_i w'_i(x) - (c_i-q_{ii})
w_i(x) + \sum_{j\neq i} q_{ij} v_j(x) = 0\quad x\in (0,b_i),
\end{equation}
with boundary conditions $w_i(0)=0, w_i'(b_i) = 1$. Since
$w_i(x)\ge0$ for $x>0$ and $w_i(0)=0$, we deduce that $w_i'(0+)\ge
0$. Furthermore, the continuity of $w_i$ and the fact that
$w_i(0)=0$ and $v_i(0)=0$ imply that
$$\sigma_i^2w_i''(0+) + 2\mu_i w_i'(0+)=0,$$ so that
$w_i''(0+)<0$, as $\mu_i>0$ by assumption.

Write now $\xi_i(x)=w_i''(x)$ for $x>0$, and denote
$\xi_i(0)=w_i''(0+)$. By twice differentiating the first equation
of the original system \eqref {2s:eq}, which is justified since
$w_i(x)\in C^4(0,b_i)$ as a consequence of the assumptions, we
find that $\xi_i(x)$ satisfies the ode
\begin{equation}
\begin{cases}
\frac{\s_i^2}{2}\xi_i''(x) + \mu_i\xi_i'(x) - (c_i -
q_{ii})\xi_i(x) +
\sum_{j\neq i}q_{ij}v''_{j}(x)=0,& x\in(0,b_i),\\
\xi_i(0)< 0, \qquad \xi_i(b_i) = 0,\qquad \xi_i(x)=0,\quad  x>b_i.\\
\end{cases}
\end{equation}
Another application of It\^o's lemma then shows that the following
representation holds true for $\xi$:
$$
\xi_i(x) = \Exp_{x}[\te{-\th_iT^i}\xi_i(X_{T^i})] +
\th_i^{-1}\sum_{j\neq i} q_{ij}\Exp_{x}\le[\int_0^{T^i}\te{-\th_i
s} v_j''(X_s)\td s\ri],
$$
where $\th_i=(c_i-q_{ii})$ and $T^i = \inf\{t\ge0:
X^i_t\notin(0,b_i)\}.$ Thus, since $\xi_i(X_{T^i})\leq 0$ and
$v_j''(x)\leq 0$, it follows that $\xi_i(x)$ is non-positive for
all $x\in(0,b_i)$ and $i\in E$. In particular, we deduce that
$x\mapsto w_i(x)$ is concave and increasing on $[0,\i)$.

Suppose now that $v\in\mc C$ and let $v_n\in\mc C\cap C^2[0,\i)$
be a sequence that pointwise increases to $v$. Then
$T_{b^v}(v)(x,i) = \lim_{n\to\i} T_{b^v}(v_n)(x,i)$, and
the concavity of $T_{b^v}(v)$ directly follows from the fact that
the pointwise limit of concave
functions is concave.\exit

We next verify that the modulated barrier strategy at $b^v$ is
optimal for the problem \eqref{eq:Uv}:

\begin{Lemma}[Optimality of barrier strategies]\label{lem:Uv2}
For $v\in\mc C$, it holds that $T_{b^v}(v)(x,i) = Uv(x,i)$ for
$x>0, i\in E$.
\end{Lemma}

\proof{\it of Lemma \ref{lem:Uv2}}: Defining again $w_i(x) = T_{b^v}(v)(x,i)$ and $f_i^v(x)$
as in \eqref{eq:ode}, we will verify that $f^v_i(x)\leq 0$ for
$x>b_i$. Next we claim that
\begin{equation}\label{eq:claim}
f^{v\prime}_i(b_i+)\leq 0,\quad\quad i\in E.
\end{equation}
The claim \eqref{eq:claim} is proved as follows. From the facts
that $w_i''(b_i)=0$ and $w_i''(x)\leq 0$ for $x<b_i$ (as a
consequence of the concavity of the $w_i$), it follows that
$w_i'''(b_i-)\geq 0$. Since both $w_i''(x)$ and $w_i'(x)$ are
continuous at $x=b_i$ and $w_i'''(b_i+)=0$, it follows by
considering the left- and right-limits of $f^{v\prime}_i(x)$ at
$x=b_i$ that $f^{v\prime}_i(b_i-)\geq f^{v\prime}_i(b_i+)$.
Finally, differentiating the identity $f^{v}_i(x)\equiv 0$ for
$x\in(0,b_i)$ shows that $f^{v\prime}_i(b_i-)=0$ and thus
\eqref{eq:claim} follows.

Noting that $f^{v\prime}_i(x) = -(c_i-q_{ii}) + \sum_{j\in E}
q_{ij}v_j'(x)$ for $x>b_i$, the concavity of $v$ together with
\eqref{eq:claim} yields then that $f^{v\prime}_i(x)\leq
f^{v\prime}_i(b_i+)\leq 0$ for $x>b_i$, which implies that
$f^{v}_i(x)\leq 0$ for $x\ge b_i$ (noting that $f^{v}_i(b_i)=0$,
by continuity).

Since $w_i$ are $C^2$ and concave and satisfy \eqref{eq:ode}, the
assertion of the Lemma follows by an argument similar to
the one used in the proof of Theorem \ref{verif:thm}. Fix an
arbitrary $D\in\mathcal D$ and let $U$ be the corresponding risk
process. Applying a generalised form of It\^{o}'s lemma to the
process $\{\te{-\Lambda_{T\wedge\tau}}w(U_{T\wedge \tau},
Z_{T\wedge \tau}), T\ge 0\}$, taking expectations and using that $f^v_i(x)\leq 0$ as in the proof of Theorem \ref{verif:thm}, we find that
$$
w(x,i) \geq  \Exp_{x,i} \left [
\te{-\Lambda_{T\wedge\tau}}w(U_{T\wedge\tau},Z_{T\wedge\tau}) +
\int_0^{T\wedge \tau\wedge\zeta} \te{-\Lambda_s} \td D_s
+1_{\{\zeta<T\wedge\tau\}} \te{-\Lambda_{\zeta}} v (U_{\zeta},
Z_{\zeta})\right ].
$$
By letting $T\to\infty$ and invoking the monotone convergence
theorem and the fact that $w$ and $v$ are non-negative and that
$f^{v}_i(x)\leq 0$, we obtain that $w(x,i) \geq  Uv(x,i)$.
Since the barrier strategy at level $b^v$ is element of $\mathcal D$
it also holds that $Uv(x,i)\geq w(x,i)$, so that
$w(x,i) = Uv(x,i)$.  \exit

The convergence of the iteration procedure is an immediate
consequence of the following contraction property of $Uv$:

\begin{Lemma}[Contraction]\label{lem:Uvn}
 The map $v\mapsto Uv$ is a contraction on $\mathcal C$
 with respect to $\|\cdot\|$.
\end{Lemma}
{\it Proof of Lemma \ref{lem:Uvn}}: Since, in view of  Lemmas
\ref{lem:Uv2} and \ref{lem:Uv}, $Uv(x,i) = \sup_b
T_b(v)(x,i) = (T_{b^v}v)(x,i)$ and $Uv\in\mc C$ for $v\in\mc C$,
it follows that for $v,w\in\mc C$
$$
\|Uv - Uw\| \leq \sup_b \|T_bv-T_bw\| \leq C \|v-w\|
$$
where $C<1$ and the second inequality follows as in the proof of
Corollary \ref{cor:ct}. Thus, $U$ is a contraction on $\mc
C$.\exit

\bigskip

\noindent{\it Proofs of Propositions \ref{prop:Uv} and \ref{prop:iter}}:
Proposition \ref{prop:Uv} directly follows by combining Lemma
\ref{lem:max} with Lemma \ref{lem:Uv2}.

From the definition of $U$ and the dynamic programming equation we
directly see that $Uv \leq V\leq Uw$ if $v\leq V\leq w$. In
particular, taking $v=v_0^-$ and $w=v_0^+$ and repeatedly applying
the former inequality yields that $v_n^-\leq V\leq v_n^+$. It
follows from Lemma \ref{lem:Uvn} that $v_n^+$ and $v_n^-$ converge
to the unique fixed point of $U$, which is therefore equal to $V$.
Next note that, in view of Lemma \ref{lem:Uv}, $v_n^\pm$ are
concave (as we took $v_0^\pm\in\mc C$), so that $V$, a pointwise
limit of concave functions, is also concave.  This completes the
proof of Proposition \ref{prop:iter}. \exit

\section{Case study: two regimes}\label{sec:casetwo}
\subsection{Positive drifts}
From now on we restrict ourselves to the case of two regimes,
$E=\{0,1\}$. For the setting of positive drifts, $\mu_0,\mu_1>0$,
we will derive a system of two non-linear equations for the
optimal dividend barriers. We will denote by $F_0$ and $F_1$ the
quadratic polynomials given by
\begin {equation} \label {Fk01}
F_k(\lambda) = \frac 12 \sigma^2_k \lambda
^2+\mu_k\lambda+q_{kk}-c_k, \quad k=0,1,
\end {equation}
with two different real roots $\lambda_1^k$ and $\lambda_2^k$.
Consider the fourth order polynomial
 \begin {equation} \label {eq:roott}
F_{0,1}(\lambda) := F_0(\lambda)F_1(\lambda)-q_{00}q_{11}.
 \end {equation}
The equation $F_k(\lambda)=0$ has two
different roots $\lambda _-^k<\lambda_+^k$ given in \eqref{noswit:root}
 and the equation $F_{0,1}(\lambda)=0$ has four real roots
  satisfying $\lambda_1<\lambda _2<0<\lambda _3<\lambda _4$.

Solving the systems of differential equation in Theorem
\ref{1equiv:thm} leads to the following result:

\begin{Prop}\label{pp2}
Suppose that $\mu_0,\mu_1>0$ and let $b_0^* < b_1^*$. Then
$(b_0,b_1)=(b^*_0,b^*_1)$ solve the two non-linear equations
\begin{align*}
q_{00}^{-1}\le[\sum \limits _{j=1}^4 \lambda_jd_jF_{0}(\lambda
_j)\te{\lambda _j b_0}\ri] &=  \frac { c_1 \le[\lambda _2^1 \te{
\lambda _1^1(b_0-b_1)} -\lambda _1^1 \te{\lambda
_2^1(b_0-b_1)}\ri] } {(\lambda _2^1-\lambda _1^1)(c_1-q_{11})}
+\frac
{q_{11}}{q_{11}-c_1}\,, \\
q_{00}^{-1}\le[\sum \limits _{j=1}^4 \lambda_j^2d_jF_{0}(\lambda
_j)\te{\lambda _j b_0}\ri] &=  \frac { \lambda _1^1\lambda _2^1
c_1 } {(\lambda _2^1-\lambda _1^1)(c_1-q_{11})}\le[ \te{ \lambda
_1^1(b_0-b_1)} - \te{\lambda _2^1(b_0-b_1)}\ri],
\end{align*}
where $d=(d_1, \ldots, d_4)'$ solves the linear system $\mathsf A
d= h$, where $h=(0,0,1,0)'$ and
\begin {equation*}
A=\left ( \begin {array}{llll}
1 & 1 & 1 & 1 \\
F_{0}(\lambda _1) & F_{0}(\lambda _2) & F_{0}(\lambda _3)
& F_{0}(\lambda _4) \\
\lambda_1\exp (\lambda_1b_{0}) & \lambda_2\exp(\lambda _2b_{0})
& \lambda_3\exp (\lambda _3b_{0}) & \lambda_4\exp (\lambda _4b_{0}) \\
\lambda _1^2\exp (\lambda_1b_{0}) & \lambda_2^2\exp(\lambda
_2b_{0}) & \lambda _3^2\exp (\lambda _3b_{0}) &\lambda _4^2 \exp
(\lambda _4b_{0})
\end {array} \right ).
\end {equation*}
\end{Prop}
The proof of Proposition \ref{pp2} is given in the Appendix.

\subsubsection{Sensitivities of the optimal barriers}\label
{nu:examples} To illustrate the effects of regime-switching and
the sensitivities of the optimal barrier levels we numerically
solved the system of non-linear equations in Propositions
\ref{pp2} for different parameter values, and compared the results
with the explicit solutions \eqref{noregO} and \eqref{noregV}
corresponding to the absence of regime-switching. The non-linear
equations were solved using a {\sf Maple} routine based on the
standard quasi-Newton method. We chose the parameters as in Table
\ref{table2} and varied $\mu_0$, $\sigma_0$, $q_{00}$ and $r_0$
individually whilst keeping the other parameters fixed ---the
results are given in Table \ref{table6}.

{\small
\begin{table}[h]
\centering
\begin {tabular}{|l||l|l|l|l|l|l|}\hline
{$i$} & {$\mu_i$} & {$\sigma_i$} & {$q_{ii}$}& {$r_i$} & $b^*_i$ &
$a_i^*$
\\\hline\hline

{0} & {0.06} & {0.24} & {-2}& {0.04} & 1.050 & 1.013
\\\hline

{1} & {0.08} & {0.30} & {-3}& {0.05} & 1.070 & 1.111
\\\hline
\end {tabular}
\caption{The parameter-set for the comparative statics.}
\label{table2}
\end{table}
}

{\small
\begin{table}
\centering
\begin {tabular}{|l||l|l|l|l|} \hline
{$\mu_0$} &{0.04} & {0.08} & 0.38 & 1.00
\\\hline\hline
{$a_0^*$} &{0.818}  &{1.100} & 0.723 & 0.169
\\\hline

{$b^*_0$} &{0.958} &{1.110} & 1.074 & 0.421
\\\hline

{$b^*_1$} &{0.974}  &{1.135} & 1.062 & 0.444
\\\hline

\end {tabular}
\begin {tabular}{|l||l|l|l|l|}\hline
{$\sigma_0$} &{0.16} &{0.20} & {0.28} &{0.32}
\\\hline\hline
{$ a_0^*$} &{0.745} &{0.896}  &{1.103} &{1.173}
\\\hline

{$b^*_0$} &{0.919} &{0.984} &{1.113} &{1.172}
\\\hline

{$b^*_1$} &{0.999} &{1.035}  &{1.104} &{1.134}
\\\hline

\end {tabular}
\smallskip

\begin {tabular}{|l||l|l|l|l|}\hline
{$q_{00}$} &{$-4$} &{$-3$}  &{$-1$} & {$-0.01$}
\\\hline\hline

{$ a_0^*$} &{1.013} &{1.013}  &{1.013} &{1.013}
\\\hline

{$b^*_0$} &{1.066} &{1.067}  &{1.036} &{1.014}
\\\hline

{$b^*_1$} &{1.082} &{1.071}  &{1.060} &{1.040}
\\\hline

\end{tabular}
\begin{tabular}{|l||l|l|l|l|}\hline
{$r_0$} &{0.02} &{0.03}  &{0.05} & {0.06}
\\\hline\hline
{$ a_0^*$} &{1.570} &{1.229}  &{0.864} &{0.753}
\\\hline

{$b^*_0$} &{1.335} &{1.174}  &{0.951} &{0.869}
\\\hline

{$b^*_1$} &{1.300} &{1.171}  &{0.989} &{0.923}
\\\hline

\end {tabular}
\caption{The optimal barriers for drifts $\mu_0$, volatilities
$\sigma_0$, transition rates $-q_{00}$ and discounting rates
$r_{0}$.}\label{table6}
\end{table}
}

We see that when the drift parameter $\mu_0$ is increased then
initially $b^*_0$ and $b^*_1$ increase, while they decrease when
the drift $\mu_0$ becomes very large. Apparently, for relatively
low drift it is optimal to reduce the probability of ruin while
for large drift the effect of discounting takes priority.
Table \ref{table6} also shows that the two barriers $b^*_0$ and
$b^*_1$ monotonically increase when $\sigma_0$ increases. A larger
volatility leads to a higher probability of ruin requiring
the company to raise the level of the barrier in order
to protect its future operations.
We can also observe the effect of the transition rates of the
underlying Markov chain. For example, if the rate is
$-q_{00}=0.01$, the chain spends a large part of the time in state
0 (in equilibrium, $3/3.01\approx 99.7\%$ of the time), which we
find back as $b^*_0=1.014$ is very close to $a_0^*=1.013$, whereas
 if $-q_{00}$ and $-q_{11}$ are of similar size, the chain spends on
average similar amounts of time in both states and the level
$b^*_0$ differs substantially from $a_0^*$.
Finally, we note that both $b^*_0$ and $b^*_1$ decrease when
the rate of discounting $r_0$ is increased: if the rate of
discounting is higher it is optimal to increase the dividend
payments by lowering the dividend barriers.

\subsection{Adverse regime-shifts: negative drift}
We next consider the case that the drift is positive in one state
and negative in the other. Intuitively it is clear that for
sufficiently small reserves a quick bankruptcy of the company is quite
likely if the drift is negative, so that it is optimal to
liquidate the company by paying out the entire reserves as a lump sum.
If, however, the negative drift is moderate
and the reserves are not too small, the expected future gains from
a regime switch to a `good' state may outweigh the effect of the
negative drift and it may be optimal to continue the business. In
that case a sensible strategy could be to liquidate the company
for small initial reserves but to pay out dividends according to a
modulated barrier strategy for larger levels of reserves, which we
formalize as follows:
\begin{Def} A {\em modulated liquidation and dividend barrier strategy}
at levels $d=(d(i), i\in E)$ and $b=(b(i), i\in E)$ is a dividend
strategy  $D^{d,b}\in\mathcal D$ satisfying
\begin{eqnarray*}
&\mrm{(i)}&\quad  \int_0^\i 1_{\{d(Z_t) < U_t^{d,b} < b(Z_t)\}}\td
D^{d,b}_t = 0, \quad\\ &\mrm{(ii)}&\quad d(Z_t) \leq U^{d,b}_t
\leq b(Z_t)\ \text{for any
$t<\tau$},\\
&\mrm{(iii)}&\quad D^{d,b}_t-D^{d,b}_{t-} = U^{d,b}_{t-}\ \text{
if $0<U^{d,b}_{t-}\leq d(Z_t)$},
\end{eqnarray*}
where $U^{d,b}$ is the insurance risk process \eqref{eq:U}
corresponding to $D^{d,b}$.
\end{Def}
Condition (iii) states that all the reserves are paid out as
dividends once the risk reserves fall below the level
$d(Z_t)$. Define next the critical levels
\begin{equation}\label{eq:deltai}
\Delta_i = \inf\le\{x\ge 0: Y_i(x)>0\ri\},
\end{equation}
where $Y_i(x):= \mu_i - c_i x + \sum_{j\neq i}q_{ij}(V_j(x) - x)$.
Note that, if $\mu_i<0$, $Y_i(x)$ is negative for all $x$ small
enough, which implies that $\Delta_i\in(0,\i]$. If $\Delta_i=+\i$,
which is the case if $\mu_i<0$ and $|\mu_i|$ is sufficiently
large, it is optimal in state $i$ to liquidate the company for any
level of the reserves, by immediately paying out all the reserves
as dividends---this can be directly checked from Theorem
\ref{verif:thm}. In the case that
$\mu_0<0<\mu_1$ and $\Delta_0<\i$ (the case
$\mu_1<0<\mu_0$ follows by relabelling the states), it turns out
that it is optimal to continue paying dividends if the reserves
are large enough, where the `liquidation' level $d^*_0>0$ solves
the smooth fit equation $V_0'(d^*_0) = 1$. The solution is
explicitly given as follows:

\begin{Prop}\label{sysnon}
Suppose that $\mu_0<0<\mu_1$ and $\D_0<\i$. (i) The optimal
strategy in \eqref{value:f} is given by the {\em modulated
liquidation and dividend barrier strategy} at levels
$d^*=(d^*_0,0)$ and $b^*=(b^*_0,b^*_1)$ that solve the system
$$
V_0'(d^*_0) = 1,\quad V_0''(b^*_0)=0,\quad V_1''(b_1^*)=0.
$$
(ii) If $b^*_1<b^*_0$, then $d^*_0,b^*_0$, and $b^*_1$ solve the
system of nonlinear equations
\begin{align*}
(q_{00}\b_1)^{-1}\sum_{j=1}^4 \l_jF_0(\l_j)
B_j(\e_{1,j}-\e_{2,j})\te{\lambda_jd_0} &= \phi + \te{(\l^1_1 +
\l^1_2)d_0}\frac{\mu_1(\lambda^1_1(\lambda^1_2)^2 - \lambda^1_2
(\lambda^1_1)^2)}{c_1 - q_{11}},\\
\sum \limits _{j=1}^4 \lambda_jB_j\te{\lambda _j b_1} &=  \frac {
c_0 \le[\lambda _2^0 \te{ \lambda _1^0(b_1-b_0)} -\lambda _1^0
\te{\lambda _2^0(b_1-b_0)}\ri] } {(\lambda _2^0-\lambda
_1^0)(c_0-q_{00})}
+\frac{q_{00}}{q_{00}-c_0}\,, \\
\sum \limits _{j=1}^4 \lambda_j^2B_j\te{\lambda _j b_1} &=  \frac
{ \lambda _1^0\lambda _2^0 c_0 } {(\lambda _2^0-\lambda
_1^0)(c_0-q_{00})}\le[ \te{ \lambda _1^0(b_1-b_0)} - \te{\lambda
_2^0(b_1-b_0)}\ri],
\end{align*}
where $\e_{i,j} = \te{\lambda_i^1d_0}[(\lambda^1_i)^2 -
\lambda_i^1\lambda_j]$, $\phi = (\lambda_1^1)^2\te{\lambda_1^1d_0}
- (\lambda_2^1)^2\te{\lambda_2^1 d_0}$, and $B=(B_1, \ldots,
B_4)'$ solves the linear system
\begin {equation}\label{mmatr2:eq}
A^* B= h,
\end {equation}
where $h=(d_0,1,q_{00},0)'$ and
\begin {equation*}
A^*=\left ( \begin {array}{llll}
\exp(\l_1 d_0) & \exp(\l_2 d_0) & \exp(\l_3 d_0) & \exp(\l_4 d_0) \\
\lambda _1\exp(\l_1 d_0) & \lambda _2\exp(\l_2 d_0) & \lambda
_3\exp(\l_3 d_0) & \lambda _4\exp(\l_4 d_0) \\
F_0(\lambda_1)\lambda_1\exp (\lambda_1b_{1}) &
F_0(\lambda_2)\lambda_2\exp(\lambda _2b_{1})
& F_0(\lambda_3)\lambda_3\exp (\lambda _3b_{1}) & F_0(\lambda_4)\lambda_4\exp (\lambda _4b_{1}) \\
F_0(\lambda_1)\lambda _1^2\exp (\lambda_1b_{1}) &
F_0(\lambda_2)\lambda_2^2\exp(\lambda _2b_{1}) &
F_0(\lambda_3)\lambda _3^2\exp (\lambda _3b_{1})
&F_0(\lambda_4)\lambda _4^2 \exp (\lambda _4b_{1})
\end {array} \right ).
\end {equation*}
The value functions are given by {
\begin {align*} 
V_0(x)&=
\begin{cases}
(1-\beta_0)\le[\alpha\frac{\te{\lambda
_1^0(x-b^*_0)}}{\lambda_1^0} + (1-\alpha)\frac{\te{\lambda
_2^0(x-b^*_0)}}{\lambda
_2^0}\ri] + \beta_0[x + \gamma], & x\in [b^*_1,b^*_0],\\
\sum\limits_{j=1}^4 B_j\te{\lambda _j x},& x\in[d^*_0, b^*_1],\\
x, & x\leq d^*_0,\\
\end{cases}\\
V_1(x)&=
\begin{cases}
q_{00}^{-1}\le[\sum \limits _{j=1}^4 B_jF_{0}(\lambda
_j)\te{\lambda _j x}\ri], & x\in[d^*_0,b^*_1],\\
\frac{\delta_2\te{\lambda_1^1x} - \delta_1\te{\lambda_2^1
x}}{(\lambda_1^1)^2\te{\lambda_1^1d^*_0} -
(\lambda_2^1)^2\te{\lambda_2^1 d^*_0}} + \beta_1 \le(x +
\frac{\mu_1}{c_1 - q_{11}}\ri), & x\in [0,d^*_0],
\end{cases}
\end {align*}
where $\alpha = \frac{\lambda _2^0}{\lambda _2^0-\lambda _1^0}$,
$\beta_i = \frac {-q_{ii}}{c_i-q_{ii}}$, and
\begin{equation*}
\gamma = q_{00}^{-1}\sum_{j=1}^4 B_jF_0(\lambda_j) \te{\lambda_j
b^*_1} - b^*_1 - \frac {\mu_0}{q_{00}-c_0}, \quad \delta_i =
\frac{\mu_1\beta_1}{c_1 -
q_{11}}(\lambda_i^1)^2\te{\lambda_i^1d^*_0} +
q_{00}^{-1}\sum_{j=1}^4 B_jF_0(\lambda_j)\lambda_j^2 \te{\lambda_j
d^*_0}.
\end{equation*}
}
\end{Prop}

Its proof is given in the Appendix. Observe that the value
function $V_0$ is not concave, as there are two disjoint intervals
where it has unit slope.

As illustration, we provide next a numerical example of a case
where a modulated liquidation-dividend strategy is optimal.

\begin{figure}[t]
\centering 
{\includegraphics[width=13cm] {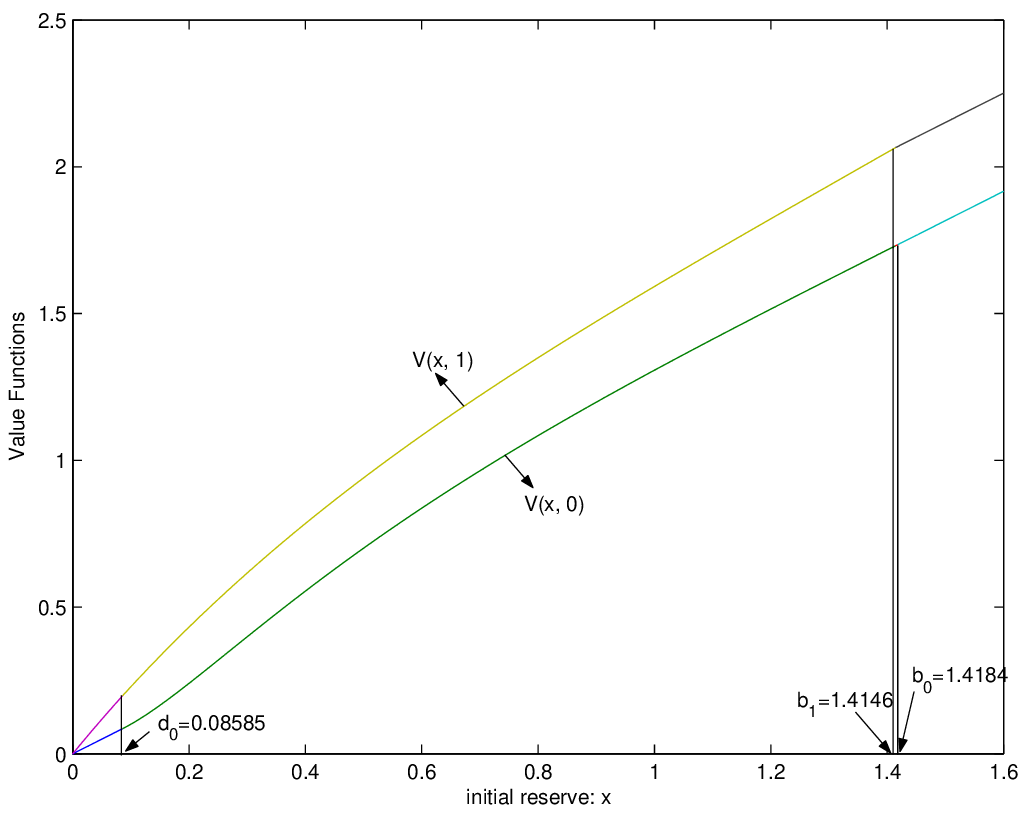}} \caption
{\label{FIG:V0}{\small The value function $V_0(x)$ and the
corresponding optimal dividend barriers $d_0, b_0, b_1$ in the
case where the parameters are $\mu_0=-0.08$,$\sigma_0=0.40$,
$q_{00}=-10$, $c_0=0.06$, $\mu_1=0.14$, $\sigma_1=0.50$,
$q_{11}=-0.001$, $c_1=0.08$. Note that $V_0$ is not concave.}}
\end{figure}

{\it Example.} Consider the case where
$\mu_0=-0.08$,$\sigma_0=0.40$, $q_{00}=-10$, $c_0=0.06$,
$\mu_1=0.14$, $\sigma_1=0.50$, $q_{11}=-0.001$, and $c_1=0.08$.
Numerically solving the system of nonlinear equations in
Proposition \ref {sysnon} we obtained  that $d_0=0.086$,
$b_0=1.418$ and $b_1=1.415$, and that value functions $V_0$ and
$V_1$ (plotted in Figure \ref{FIG:V0}) are given as follows: {
$$
 V_0(x)=
\begin {cases}
x+0.317,  &  x\ge 1.418; \\
-1174e^{-10.725x} +0.994x +2.43\cdot10^{-4}
e^{11.725(x-1.418)}+0.325\,,
 & 1.415\leq x \leq 1.418;\\
0.266e^{-10.725x}-1.252e^{-1.536x} + 1.039e^{0.417x}\,, &0.086<x<1.415;\\
x\,, & 0\leq x \leq 0.086.
\end {cases}
$$
and
\begin {equation*}
 V_1(x)=
\begin {cases}
x+0.651,  &  x\ge1.415; \\
-2.04\cdot 10^{-5} e^{-10.725x} -1.220e^{-1.536x}
+1.222e^{0.417x}\,, & 0.086 \leq x \leq 1.415;\\
-1.230e^{-1.541x}+1.209e^{0.421x}+0.012x+0.021\,, & 0\leq x \leq
0.086.
\end {cases}
\end {equation*}
}

\section {Conclusion}\label{sec:concl}

In this paper we have shown that, in the presence of
regime-shifts, the optimal dividend policy is given by a threshold
strategy set at a level that is a function of the current
regime. That is to say, the policy that maximises the expectation of
the net present value of the paid dividends until the moment of
default consists of paying out as dividends the overflow of the
cash reserves above a certain optimal threshold, where this
threshold jumps up or down exactly at the moment when
the regime shifts. Hence, at the moment of a
regime shift, when the key parameters such as drift, volatility
and discounting may change, it may be optimal to make a lump-sum dividend
payment, namely when the threshold level jumps below the current
level of the cash reserves. We presented a contraction algorithm
for the computation of the optimal threshold levels.
As
a case study we numerically investigated the parameter
sensitivities of the levels in the case of two regimes.
It would be desirable to systematically explore the dependence
of the optimal threshold levels on key parameters, and its
financial significance, which could be achieved by an
analytical investigation of its form in
specific parametric models; this is a topic left for future
research.




\newpage
\appendix
\centerline{\bf\large Appendix}
\section{Proofs}

\subsection{Proof of the bounds (Proposition \ref{prop:apriori})}
To prove the upper and lower bounds in \eqref{lub} we consider two
auxiliary optimal switching problems where not only the dividend
payout but also the regime is a control variable. An admissible
switching strategy $\s=\{Z^\s_t, t\ge 0\}$ is an $\mbf F$-adapted
$E$-value process that indicates the current regime. The two
control problems are then given by
\begin{equation}
v^+(x) := \sup_{\s\in\mc S, D\in \mc D} \Exp_x\le[\int_0^{\tau^\s}
\te{-\Lambda^\s_s}\td D_s\ri], \qquad v^-(x) := \inf_{\s\in\mc
S}\Exp_x\le[\int_0^{\tau^\s} \te{-\Lambda^\s_s}\td D_s^{{-}}\ri],
\end{equation}
where $D^{-}$ denotes the constant barrier strategy at level
$b_{-}$ (where $b_-$ denote the optimal barriers corresponding
to $V_-$), $\mc S$ and $\mc D$ are the sets of all admissible
switching and dividend strategies, $\L^\s_s=\int_0^s c(Z_u^\s)\td
u$ and $\tau^\s$ is the corresponding ruin time. As the
regime-switching process $Z$ is one particular admissible
switching strategy, the upper and lower bounds in \eqref{lub} will
follow once we have shown that $v^+\leq V_+$ and $v^-\geq V_{-}$.

In the proof we will use the following sub- and super-harmonicity
properties:

\begin{Lemma}\label{lem:apriori}
For all $i\in E$, it holds that
\begin{eqnarray}\label{eq:gen*1}
\mc G_i V_{+}(x) - c_i V_{+}(x) &\leq& 0\q \text{for all $x>0$},\\
\mc G_i V_{-}(x) - c_i V_{-}(x) &\geq& 0 \q\text{for all
$x\in(0,b_{-})$}, \label{eq:gen*2}
\end{eqnarray}
where $\mc G_i$ is the infinitesimal generator of $X^i_t=x + \mu_i
t + \sigma_i W_t$.
\end{Lemma}
\proof Since $V_{+}$ is the value function corresponding to the
optimal dividend problem without regime-switching, and with
volatility, drift and discounting $1, \max_{i\in
E}\frac{\mu_i}{\sigma^2_i}, \min_{i\in E}\frac{c_i}{\sigma^2_i}$,
it solves a corresponding HJB equation. In particular, $V_+'$ and
$V_{+}$ are both positive, so that, in view of the form of the
drift and discounting it follows that \eqref{eq:gen*1} holds true.
By a similar argument it can be verified that \eqref{eq:gen*2}
holds true.\exit

\proof{\it of Proposition \ref{prop:apriori}.} Fixing an arbitrary
admissible switching and dividend strategies $\s$ and  $D$ and
denoting by $U^{\s,D}$ the corresponding risk process, an
application of It\^o's lemma shows that
\begin{eqnarray*}
  \te{-\Lambda^\s_{t\wedge\tau^{\s}}} V_{+}(U^{\sigma,
  D}_{t\wedge\tau^{\s}}) &=& V_{+}(x) + \int_0^{t\wedge\tau^{\s}}
  \sum_{i\in E}1_{\{Z^\s_s=i\}}
  \te{-\Lambda^\s_s}(\mc G_i - c_i)V_{+}(U_s^{\s,D})\td s\\
&+& \sum_{0\leq s\leq
t\wedge\tau^{\s}}\te{-\L^{\s}_s}\le[V_{+}(U_{s-}^{\s,D} - \D D_s)
- V_{+}(U_{s-}^{\s,D}) + V_+^{\prime}(U_{s-}^{\sigma,
  D})\D D_s\ri]1_{\{\D D_s > 0\}} \\
  &+&
  M^\s_{t\wedge\tau^{\s}} - \int_{0}^{t\wedge\tau^{\s}}
  \te{-\Lambda^\s_s} V_+^{\prime}(U_{s-}^{\sigma,
  D})\td D_s,
\end{eqnarray*}
where $M^{\sigma}$ is some local martingale which is a
supermartingale as it is bounded below. Taking note of Lemma
\ref{lem:apriori} and the facts that $V_+^{\prime}\ge 1$ and
$V_+(0)=0$, it follows by rearranging and taking expectations that
\begin{equation}\label{eq:V**}
V_{+}(x) \ge \Exp_x\le[\int_0^{\tau^\s} \te{-\Lambda^\s_s}\td
D_s\ri].
\end{equation}
Subsequently taking the supremum in \eqref{eq:V**} over all
$\sigma\in\mc S$ and $D\in\mc D$ shows that $V_+(x)\ge v^+(x)$. By
a similar line of reasoning it can be verified that $V_{-}(x) \leq
v^-(x)$ for all $x\leq b_{-}$. In particular, writing $\c=(b_{-},
\ldots, b_{-})$ it follows that, for all $x\leq b_{-}$,
\begin{equation}\label{eq:VVV}
V_{-}(x) \leq V^{\c}(x,i).
\end{equation}
Observing that, for $x\ge b_{-}$, $V_{-}'(x)=1$ whereas
$V^{\c\prime}_i(x)\ge 1$, we see that Eqn. \eqref{eq:VVV} is valid
for all $x\ge 0$, and, since $V^{\c}\leq V$, the proof of
\eqref{lub} is complete.\exit

\subsection{The dynamic programming equation (Proposition \ref{prop:DP})}
The proof is an adaptation of a classical line of reasoning to a
regime-switching setting. We start with the following two lemmas:
\begin{Lemma} For $x\ge y \ge 0$ and $i\in E$ it holds that
$$x-y \leq V(x,i) - V(y,i) \leq \le(1 -
\frac{W^{(\theta_i)}(y)}{W^{(\theta_i)}(x)}\ri)V(x,i),$$ where
$\theta_i=c_i-q_{ii}$ In particular, it follows that $V(\cdot,i)$
is Lipschitz continuous.
\end{Lemma}
\proof Let $\e>0$ and let $D(u,i)$ be an $\e$-optimal strategy for
$U_0=u, Z_0=i$, and consider the strategies $D'_t(u,y) =
(u-y)\mathbf 1_{\{t=0\}} + D_t(y,i)\mathbf 1_{\{t>0\}}$ (``pay a
lump sum $u-y$ and follow then the strategy $D(y,i)$'') and $\T
D_t(u,x)=\mathbf 1_{\{t>\t(x), Z_{\t(x)}=i\}}D(x,i)$ for $x\ge
u\ge y$ (``wait until the first time $\t(x)$ that the reserves
reach the level $x$; if no regime-switch has occurred by then,
follow the strategy $D(x,i)$, otherwise don't pay any
dividends''). Then it follows that
\begin{eqnarray*}
V(x,i) &\ge& V_{D'(x,y)}(x,i) \ge x-y + V_{D(y,i)}(y,i) \ge x-y +
V(y,i) - \e, \\
V(y,i) &\ge& V_{\T D(y,x)}(y,i) \ge t_i(y,x) V_{D(x,i)}(x,i) \ge
t_i(y,x)(V(x,i) - \e),
\end{eqnarray*}
where $t_i(y,x) = E_y[\te{-\th_i \t(x)}\mathbf 1_{\{\t(x)<\t\}}] =
\frac{W^{(\theta_i)}(y)}{W^{(\theta_i)}(x)}$. Letting $\e\to0$ the
bounds follow.\exit
\begin{Lemma}
  Let $M>0, \epsilon>0$. There exists a $\WT D\in\mathcal D$ such
  that
  $$
\max_{i\in E}\sup_{x\in[0,M]}(V(x,i) - V_{\WT D}(x,i))<\e.
  $$
\end{Lemma}
\proof Choose a grid $(x_{(j)}:= \frac{jM}{N}, j=0, \ldots, N)$ of
$[0,M]$, where $N< \e^{-1}$ is chosen such that $\max_{i\in
E}m_{V,i}(N^{-1})< \e$, with $m_{V,i}$ the modulus of continuity
of $V(\cdot,i)$
$$m_{V,i}(h)=\sup_{|x-y|<h,x,y\in[0,M]}|V(x,i)-V(y,i)|.$$
Let $D^{i,j}$ be $\e$-optimal strategies corresponding to
$U_0=x_{(j)}$ and $Z_0=i$, that is, $V(x_{(j)},i) -
V_{D^{i,j}}(x_{(j)},i) < \e$, and define the strategy $\WT D$
depending on $U_0=x$ and $Z_0=i$ as 'pay a lump sum
$(x-x_{(j^*)})$ and follow then the strategy $D^{i,j^*}$' where
$j^*=\max\{j:x_{(j)}\leq x\}$:
$$
\WT D_t = (x-x_{(j^*)})\mathbf 1_{\{t=0\}} + D^{i,j^*}_t\mathbf
1_{\{t\ge 0\}}.
$$
Then it follows that
\begin{eqnarray*}
|V(x,i) - V_{\WT D}(x,i)| &\leq& |V(x,i)-V(x_{(j^*)},i)| +
|V(x_{(j^*)},i)-V_{D^{i,j^*}}(x_{(j^*)},i)|\\ &+&
|V_{D^{i,j^*}}(x_{(j^*)},i) - V_{D^{i,j^*}}(x,i)|\leq \e + \e + \e
= 3\e.
\end{eqnarray*}
As this estimate holds for arbitrary $x\ge 0$ and $i\in E$, the
proof is complete.\exit

\proof{\it of Proposition \ref{prop:DP}}: Denote by $w$ the right-hand
 side of \eqref{eq:DP} and by $D\in\mc D$ and $U$ an arbitrary
  admissible strategy and the corresponding cash reserves.
To show that $V\leq w$, we verify that $V_D\leq
w$:
\begin{eqnarray*}
V_D(x,i) &=&
\Exp_{x,i}\le[\int_0^{\tau\wedge\zeta}\te{-\Lambda_t}\td D_t +
\mathbf 1_{\{\zeta<\tau\}}
\Exp_{x,i}\le[\int_{\zeta}^\tau \te{-\Lambda_t}\td D_t\bigg|\mc F_{\zeta}\ri]\ri]\\
&\leq& \Exp_{x,i}\le[\int_0^{\tau\wedge\zeta}\te{-\Lambda_t}\td
D_t + \mathbf 1_{\{\zeta<\tau\}}\te{-\Lambda_{\zeta}}V(U_{\zeta},
Z_{\zeta})\ri]\leq w(x,i).
\end{eqnarray*}
To prove the opposite bound $w\leq V$ we will show that, for given
$\e>0$ and $D\in\mc D$, there exists a strategy $D(\e)\in\mc D$
such that $w_D\leq V_{D(\e)} + \text{const}\cdot\e$, where $w_D$
denotes the expectation in \eqref{eq:DP}. Fixing $M>0$ such that
$P_{x,i}(X_\zeta>M)< \e$ for all $i\in E$, we denote by
$D^\e\in\mc D$ a dividend strategy that pays out $x-M$, if
$U_0=x>M$, and that is $\e$-optimal, uniformly over starting
values $(i,x)\in E\times[0,M]$, that is, $V(x,i)< V_{D^\e}(x,i) +
\e$ for all $(i,x)$ in this set. 
Letting $\theta$ denote the shift-operator, note that 
$D(\e):=D_t\mathbf
1_{\{t<\zeta\}} 
+ D^\e_{t-\zeta}\circ\theta_{\zeta}\mathbf 1_{\{t\ge\zeta\}}\in \mc
D$, and that it satisfies
\begin{eqnarray*}
\lefteqn{\Exp_{x,i}\le[\int_0^{\tau\wedge\zeta}\te{-\Lambda_t}\td D_t + \te{-\Lambda_{\tau\wedge\zeta}}
V(U_{{\tau\wedge\zeta}}, Z_{{\tau\wedge\zeta}})\ri]}\\
&\leq& \Exp_{x,i}\le[\int_0^{\tau\wedge\zeta}\te{-\Lambda_t}\td D_t + \te{-\Lambda_{\tau\wedge\zeta}}
[V_{D^\e}(U_{{\tau\wedge\zeta}}, Z_{{\tau\wedge\zeta}})+\epsilon + C1_{\{U_{\tau\wedge\zeta}>M\}}]\ri]\\
&\leq& V_{D(\e)}(x,i) + \e + CP_{x,i}(X_{\zeta}>M) \leq
V(x,i) + \e(1 + C),
\end{eqnarray*}
where $C=\max_i (V(x,i)-x)$, which is finite in view of Proposition \ref{prop:apriori}.\exit

\subsection{The optimal levels (Lemma \ref{lem:max})}
\proof{\it of Lemma \ref{lem:max}}: Since $A^v_i(b)$ is continuous, it attains its maximum at
some $b\in[0,\i]$. By straightforward calculus it can be verified
that its derivative is given by
$$
W^{(q)\prime}_i(b) A^{v\prime}_i(b) = -
\frac{W^{(q)\prime\prime}_i(b)}{W^{(q)\prime}_i(b)} + \int_0^b
\sum_{j\neq i}\frac{q_{ij}}{\theta_i}v'_j(y)k_i(b-y,b)\td y
$$
where $k_i(y,b) = W^{(q)\prime}_i(y) - W_i^{(q)}(y)\ell^{(q)}_i(b)$
with $\ell^{(q)}_i(b)=W^{(q)\prime\prime}_i(b)/W^{(q)\prime}_i(b)$
and $q=\theta_i$. From eq. \eqref{eq:Wq} it is straightforward to
check  that $\ell^{(q)}_i(b)$ converges to $\l_i^+(q)$ as
$b\to\i$, and that $k_i(y,b) \leq 2{\s^{-2}_i}\te{\l^-_i(q) y}$.
By dominated convergence, it then follows that
$W^{(q)\prime}_i(b)A^{v\prime}_i(b)$ tends to $-\l^+_i(q) <0$ as
$b\to\i$. Since it also holds that $A^{v\prime}_i(0+) =
4\mu_i/\sigma^4_i > 0$, we see that $A^v_i(b)$ attains its maximum
on $(0,\i)$. Therefore $A_i^{v\prime}(b_i)=0$ which
implies that $(T_bv)^{\prime\prime}_i(b_i)
= 0$, in view of the definition of $T_bv$. Since $T_bv''(x,i)=0$
for $x>b_i$, it follows that
$T_{b^v}(\cdot,i)\in C^2[0,\infty)$
for $i\in E$. \exit

\subsection{The case of two regimes (Propositions \ref{pp2} and \ref{sysnon})}

\begin {Lemma}\label{lee00}
If there exists $0<b_\imath<b_\jmath$ such that system~\eqref
{2s:eq} holds true, then
\begin{eqnarray*}
V(x,\jmath)&=& \frac {\lambda _2^\jmath c_\jmath\exp [\lambda
_1^\jmath(x-b_\jmath)]}{\lambda _1^\jmath(\lambda
_2^\jmath-\lambda _1^\jmath)(c_\jmath-q_{\jmath\jmath})}+\frac
{\lambda _1^\jmath c_\jmath\exp [\lambda
_2^\jmath(x-b_\jmath)]}{\lambda _2^\jmath(\lambda
_1^\jmath-\lambda _2^\jmath)(c_\jmath-q_{\jmath\jmath})}
+\frac{q_{\jmath\jmath}x}{q_{\jmath\jmath}-c_\jmath} \nonumber \\
&&+\frac{q_{\jmath\jmath}(q_{\jmath\jmath}-c_\jmath)(V(b_\imath,\imath)-b_\imath)-\mu_\jmath
q_{\jmath\jmath}}{(q_{\jmath\jmath}-c_\jmath)^2}\,,\qquad x\in
(b_\imath,b_\jmath).
\end{eqnarray*}
\end{Lemma}

\proof When $0<b_\imath<x<b_\jmath$, it follows immediately from
system~\eqref{2s:eq} that
$$
\frac 12 \sigma^2_\jmath V''(x,\jmath)+\mu_\jmath V'(x,\jmath)
+(q_{\jmath\jmath}-c_\jmath)V(x,\jmath)=q_{\jmath\jmath}(x-b_\imath+V(b_\imath,\imath)),
$$
whose general solution is of the following form
$$
V(x,\jmath)=k_1 \exp {(\lambda _1^\jmath x)}+k_2 \exp{(\lambda
_2^\jmath x)}+k_3x+k_4
$$
for any $k_1,\,k_2 \in \mathbb R$ since the quadratic
characteristic equation $F_\jmath(\lambda)=0$ of its corresponding
homogeneous equation has two roots $\lambda _1^\jmath$ and
$\lambda _2^\jmath$ and its particular solution is obviously
$k_3x+k_4$, where
\begin {equation}
k_3=\frac {q_{\jmath\jmath}} {q_{\jmath\jmath}-c_\jmath}\quad
\text {and}\quad k_4=\frac
{q_{\jmath\jmath}(q_{\jmath\jmath}-c_\jmath)(V(b_\imath,\imath)-b_\imath)-\mu_\jmath
q_{\jmath\jmath}}{(q_{\jmath\jmath}-c_\jmath)^2}\,.
\end {equation}
Using boundary conditions $\frac {\partial V(x,\jmath)} {\partial
x}|_{x=b_\jmath} =1$ and $\frac {\partial^2 V(x,\jmath)}
{\partial^2 x}|_{x=b_\jmath}=0$ then yields
\begin {equation}
k_1=\frac {\lambda _2^\jmath c_\jmath\exp (-\lambda _1^\jmath
b_\jmath)}{\lambda _1^\jmath(\lambda _2^\jmath-\lambda
_1^\jmath)(c_\jmath-q_{\jmath\jmath})} \quad \text {and}\quad
k_2=\frac {\lambda _1^\jmath c_\jmath\exp (-\lambda _2^\jmath
b_\jmath)}{\lambda _2^\jmath(\lambda _1^\jmath-\lambda
_2^\jmath)(c_\jmath-q_{\jmath})}\,.
\end {equation}
The proof is complete. $\hfill \Box$

\proof{\it of Proposition \ref{pp2}.} In view of Lemma \ref{lee00}
to complete the proof it remains to derive the system. For
$0<x<b_{\imath}<b_\jmath$, it follows from~\eqref{2s:eq} that
$V(x,\imath)$ satisfies a four-order linear homogeneous ordinary
differential equation with the characteristic equation
$F_0(\lambda)F_1(\lambda)-q_{00}q_{11}=0$ having four real roots
$\lambda_1<\lambda_2<\lambda_3<\lambda_4$. Thus, for
$0<x<b_\imath$, $V(x,\imath)$ and $V(x,\jmath)$ can be
respectively expressed as
\begin{eqnarray}
V(x,\imath)&=&\sum\limits_{j=1}^4 d_j\exp (\lambda _j
x), \label {istate}\\
V(x,\jmath)&=&q_{\imath\imath}^{-1}[\sum \limits _{j=1}^4
d_jF_{\imath}(\lambda _j)\exp (\lambda _j x)], \label {jstate}
\end{eqnarray}
where coefficients $d_i, i=1,\ldots, 4$ are to be determined. By
expressing the boundary conditions
$$V(0,\imath)=V(0,\jmath)=\le.\frac {\partial^2
V(x,\imath)} {\partial^2 x}\ri|_{x=b_{\imath}}=0\q\text{ and }
\le.\frac{\partial V(x,\imath)} {\partial x}\ri|_{x=b_{\imath}}=1
$$
in terms of these coefficients we arrive at the matrix equation
$\mathsf A d = h$. The equations for the optimal levels follow
since $V(x,\jmath)$ is $C^2$ at $b_\imath$ (noting that any two of
the three equations implies the third one). \exit

The system in Proposition \ref{pp2} can be solved explicitly as
shown explicitly in the following result. To that end we introduce
the functions $\tilde g_{\imath,k}$ and $g_{\imath,k}$, $k=1,2$,
as follows
\begin{eqnarray*}
\tilde g_{\imath,k}(x) &=& e^{\lambda_k x} +C_k e^{\lambda_3 x} -
(C_k+1) e^{\lambda_4 x}\\
 g_{\imath,k}(x)&=& F_\imath
(\lambda_k)e^{\lambda_k x} + C_k F_\imath (\lambda_3) e^{\lambda_3
x}- (C_k+1) F_\imath (\lambda_4) e^{\lambda_4 x}
\end{eqnarray*}
where $C_k = \frac {F_\imath(\lambda_k)-F_\imath
(\lambda_4)}{F_\imath(\lambda_4)-F_\imath(\lambda_3)}$ and we
write
$$
G = \tilde g'_{\imath,1}(b_\imath)\tilde g''_{\imath,2}(b_\imath)-
\tilde g'_{\imath,2}(b_\imath)\tilde g''_{\imath,1}(b_\imath).
$$

\begin{Lemma} \label{lee22}
If $G\neq 0$ it holds for $0 \leq x<b_\imath$ that
\begin{eqnarray}
V(x,\imath)&=& G^{-1}[\tilde g''_{\imath,2}(b_\imath)\tilde
g_{\imath,1}(x)-\tilde g''_{\imath,1}(b_\imath)\tilde
g_{\imath,2}(x)]  \label {value:i}\\
V(x,\jmath)&=& (q_{\imath\imath}G)^{-1}[\tilde
g''_{\imath,2}(b_\imath)g_{\imath,1}(x)-\tilde
g''_{\imath,1}(b_\imath)g_{\imath,2}(x)]\label{value:j}
\end{eqnarray}
\end{Lemma}
\proof From the first two equations of the system $\mathsf A d =
h$ and in view of $\lambda_4>\lambda_3>0$ and $\mu_{\imath}>0$, we
can express coefficients $d_3$ and $d_4$ in terms of $d_1$ and
$d_2$ as follows $$ \left \{
  \begin {array}{ccc}
 d_3&=&\frac {(F_{\imath}(\lambda_1)-F_{\imath}(\lambda_4))d_1
     + (F_{\imath}(\lambda_2)-F_{\imath}(\lambda_4))d_2}
     {F_{\imath}(\lambda_4)-F_{\imath}(\lambda_3)}\\
\\
 d_4&=&\frac {(F_{\imath}(\lambda_3)-F_{\imath}(\lambda_1))d_1
     + (F_{\imath}(\lambda_3)-F_{\imath}(\lambda_2))d_2}
     {F_{\imath}(\lambda_4)-F_{\imath}(\lambda_3)}
  \end {array} \right.
$$
Substituting this into~\eqref {istate} and~\eqref {jstate} yields
\begin {eqnarray}
V(x,\imath)&=&d_1 \tilde g_{\imath,1}(x)+d_2\tilde
g_{\imath,2}(x), \qquad\quad 0<x<b_{\imath}.\label {inewstate}\\
V(x,\jmath)&=&q_{\imath\imath}^{-1}[d_1g_{\imath,1}(x)+
d_2g_{\imath,2}(x)], \qquad 0<x<b_{\imath}.\label {jnewstate}
\end {eqnarray}
The last two equations of $\mathsf A d = h$ then can be rewritten
as
\begin {equation*}
\tilde g'_{\imath,1}(b_\imath)d_1+\tilde
g'_{\imath,2}(b_\imath)d_2=1 \quad \text{and} \quad \tilde
g''_{\imath,1}(b_\imath)d_1+\tilde g''_{\imath,2}(b_\imath)d_2=0
\end {equation*}
with a unique solution
\begin {equation} \label {l12}
(d_1,d_2)=G^{-1} (\tilde g''_{\imath,2}(b_\imath), -\tilde
g''_{\imath,1}(b_\imath))
\end {equation}
according to Cram\'{e}r's rule (as $G\neq0$).\exit

\proof{\it of Proposition \ref{sysnon}.} The structure of the
proof is analogous to that of Theorem \ref{1equiv:thm}. As the
value function $V_0$ will not be concave some parts of the proof
has to be modified. The steps are outlined as follows:

1. The value function of a modulated liquidation-dividend strategy
$V^{d,b}$ satisfies $V = \tilde T_{d,b} V$ where
\begin{equation}\label{eq:ttt}
\tilde T_{d,b}(f)(x,i) =
\begin{cases}
  x, & x\in [0,d_i];\\
W_i^{(\theta_i)}(x-d_i) \T A^{f}(d_i,b_i) + f_{1-i}(d_i)
Z_i^{(\theta_i)}(x-d_i) \\
\quad - \frac{q_{i,1-i}}{\theta_i}
\int_0^{x-d_i}f_{1-i}(y+d_i)W^{(\theta_i)}_i(x-d_i-y)\td y, &
x\in[d_i,b_i];\\
x - b_i + \tilde T_{d,b}f(x,i), & x\ge b_i,\\
\end{cases}
\end{equation}
where $\theta_i=c_i - q_{ii}$, $Z_i^{(\theta_i)}(x)=1+\theta_i
\int_0^x W^{(\theta_i)}_i(y)\td y$, and
$$\T A^{f}(d_i,b_i) = \frac{1 - \theta_i
W^{(\theta_i)}_i(b_i-d_i) +
\frac{q_{i,1-i}}{\theta_i}\int_0^{b_i-d_i}f_{1-i}(y-d_i)W^{(\theta_i)\prime}_i(b_i-d_i-y)\td y}%
{W^{(\theta_i)\prime}_i(b_i-d_i)}.
$$

2. The map $f\mapsto \sup_{d,b}\tilde T_{d,b}f$ is a contraction
on $\mc B$. As a consequence, there exists a function $w$ with
$w=\sup_{d,b}\tilde T_{d,b}w$. By similar arguments as Lemma
\ref{lem:max} it can be verified that there exist $d_0,b_0,b_1$
with $0< d_0 < b_0, b_1$ such that
$$w_0'(d_0)=1,\quad w_i''(b_i)=0,
\quad i=0,1.$$
Let now $b_\imath=\min (b_i, b_{1-i})$ and $b_\jmath=\max (b_i,
b_{1-i})$, and define
$$
\WT w(x,\jmath) = \begin{cases} w(x,\jmath), & x\leq \gamma;\\
x-\gamma + w(\gamma,\jmath), & x>\gamma,
\end{cases}
$$
where $\gamma = \inf\{x\in(b_\imath,b_\jmath]:
f_{\imath}''(x)=0\}$ (with $\inf\emptyset=b_\jmath$), and
$f_{\imath}(x) =
\mu_{\imath}-(c_{\imath}+q_{\imath\jmath})w_{\imath}(x) +
q_{\imath\jmath}w_\jmath(x)$. We will directly verify that
\begin{equation}\label{eq:negclaim}
\text{$f_j'(x)\leq 0$ for all $x> b_j$, $j=0,1$,}
\end{equation}
which implies that $f_j(x)\leq 0$ for $x>b_j$, since
$f_j(b_j)=0$).

Indeed, note that, for $x>\gamma$, $f_j'(x)=-c_j (j=0,1)$.
Further, if $\gamma>b_{\imath}$, it is straightforward to verify
from the equations satisfied by $w$ that, for $b_{\imath}< x <
\gamma $, $w''_{\jmath}(x)= A\lambda_1\te{\lambda_1 x} +
B\lambda_2\te{\lambda_2 x}$ for some $A,B>0$ and
$\lambda_1<0<\lambda_2$, which implies that $w''(x)<0$ for
$b_{\imath}< x < \gamma $, as $w''(\gamma)=0$. By an argument as
in Lemma \ref{lem:Uv2}, it then follows that \eqref{eq:negclaim}
also holds true for $b_{\imath}<x<\gamma$.

3.  In particular, there exist levels $0<d_0<b_0,b_1$ for which
$V$ solves the system
\begin {equation}\label{3s:eq}
\begin {cases}
 \frac 12
{\sigma_i}^2V^{\prime\prime}_i(x)+\mu_iV^\prime_i
(x)-(q_{i,1-i}+c_i)V_i(x)=-q_{i,1-i}V_{1-i}(x),        & d_i<x<b_i; \\
V_i(x) = x,                                      & 0\leq x\leq
d_i;\\
V_i(x)= x-b_i+V_i(b_i),                          & x\ge b_i;\\
V_i(0)= 0, V_0'(d_0) = 1, V'_i(b_i) = 1, V_i''(b_i)=0, &
\end {cases}
\end {equation}
for $i=0,1$, $d_1=0$, and $V_i(x)=V(x,i)$.

Reasoning as in the proof of Proposition \ref{pp2} we can
subsequently derive the expressions in Proposition \ref{sysnon}.


\end{document}